\documentstyle[sprocl]{article}

\def\sst{\scriptscriptstyle}

\def\be{\begin{equation}}
\def\ee{\end{equation}}
\def\bea{\begin{eqnarray}}
\def\eea{\end{eqnarray}}
\let\a=\alpha 
\let\b=\beta 
\let\g=\gamma  
\let\d=\delta 
\let\e=\epsilon
 
\let\h=\eta  
\let\k=\kappa
\let\l=\lambda
\let\m=\mu
\let\n=\nu 
 
\let\r=\rho
\let\s=\sigma 
\let\t=\theta

\let\la=\label

\let\bm=\bibitem

\def\nn{\nonumber} 
\def\bd{\begin{document}} 
\def\ed{\end{document}}
\def\ds{\documentstyle} 
\def\v{\wedge}
\def\ba{\begin{array}}
\def\ea{\end{array}}
\def\ft#1#2{{\textstyle{{\scriptstyle #1}\over {\scriptstyle #2}}}}
\def\fft#1#2{{#1 \over #2}}
\def\sst#1{{\scriptscriptstyle #1}}
\def\oneone{\rlap 1\mkern4mu{\rm l}}
\def\[{{[}}
\def\]{{]}}
\def\es{\vspace{10pt}}
\def\hm{{\hat\mu}} 
\def\hn{{\hat\nu}} 
\def\4{\m_1\cdots \m_4}
\def\5{\m_1\cdots \m_5}
\def\6{\m_1\cdots \m_6}\def\h6{\hm_1\cdots \hm_6}
\def\Hat#1{\widehat{#1}}
\def\qq{\quad\quad}
\def\uC{{\underline C}}

\newcommand{\ho}[1]{$\, ^{#1}$}
\newcommand{\hoch}[1]{$\, ^{#1}$}
\newcommand{\Str}{\rm Str\, }
\newcommand{\p}{\partial}
\newcommand{\tb}{\bar\theta}
\newcommand{\cm}{{\cal M}}
\newcommand{\ha}{{\hat A}}
\newcommand{\hb}{{\hat B}}
\newcommand{\hc}{{\hat C}}
\newcommand{\ham}{{\hat M}}
\newcommand{\eq}[1]{(\ref{#1})}

\def\ii{\.I}
\def\o{\"o}
\def\uu{\"u}
\def\wi{Wigner-\.In\"on\"u} 

\begin{document}

\hfill{CTP TAMU-78/96}

\hfill{hep-th/9612220}

\hfill{December 1996}

\vspace{30pt}

\centerline{\Large \bf Chiral Reductions of the M-Algebra}

\vspace{30pt}

\centerline{\large Ergin Sezgin\footnote{Research supported in part by 
NSF Grant  PHY-9411543}}

\vspace{15pt}

\centerline{\it Center for Theoretical Physics, Texas A\&M University,}
\centerline{\it College Station, Texas 77843, U.S.A.}

\vspace{50pt}

\centerline{ABSTRACT}
\bigskip

\abstracts{ We present the chiral truncation of the eleven
dimensional M-algebra down to ten and six dimensions. In ten dimensions,
we obtain a topological extension of the $(1,0)$ Poincar\'e superalgebra
that includes super one-form and super five-form charges. Closed super
three- and seven-forms associated with this algebra are constructed. In
six dimensions, we obtain a topological extension of the $(2,0)$ and
$(1,0)$ Poincar\'e superalgebras. The former includes a quintet of super
one-form charges, and a decuplet of super three-form charges, while the
latter includes a triplet of super three-form charges.}
\vspace{2cm}
\centerline{\it DEDICATED TO THE MEMORY OF ABDUS SALAM}

\vfill\eject

It is well known that Poincar\'e superalgebras are modified in presence
of $p$-brane solitons \cite{p,pkt1}. An interesting case to consider is
the modification of the eleven dimensional Poincar\'e superalgebra due
to the existence of the supermembrane \cite{ds} and superfivebrane solitons
\cite{guven}. Recently we have obtained such a modification, which we
have called the M-algebra \cite{es1}. This algebra contains a number of
subalgebras that have been obtained previously \cite{df1,g,bs1,bs2,es}.
Due to lack of space, we shall not review the motivations for the
M-algebra, and the assumptions we have made to obtain it. Suffice it
to mention, however, that it contains super one-, two- and five-form
charges. In particular, the latter two are associated with the
supermembrane and superfivebrane solitons of M-theory. 

The purpose of this note is to construct the analog of the M-algebra
algebra in ten and six dimensions, where chiral $(p,q)$ superalgebras
exist, i.e. those which admit independent $p$ left- and $q$ right-handed
supercharges, with $p\ne q$. Assuming Lorentzian signature, chiral
superalgebras exist in $D=2,6,10$. However, the two dimensional target
space is rather special and requires a separate treatment. Here, we
shall concentrate on six and ten dimensions. Furthermore, we will
discuss mostly the $(1,0)$ and $(2,0)$ symmetries. The remaining case of
$(2,1)$ symmetry in six dimensions is also special, in that we are not
aware of any field theoretic realizations for it. We shall not consider 
this case any further in this note.

The construction of the M-type algebras in ten and six dimensions will
be achieved by dimensional reduction of the eleven dimensional
M-algebra, followed by suitable chiral truncations. To establish our
notation, let us begin by defining the the Poincar\'e supercharges
$Q_A=(P_\m, Q_\a)$ which generate supertranslations in flat $D=11$
superspace parametrized by the coordinates $Z^M=(X^\m,\t^\a)$,
$\m=0,1,...,10$, where $\t^\a$ are anticommuting 32 component Majorana
spinors. The M-algebra contains the additional charges $Z^A$, $Z^{AB}$
and $Z^{A_1\cdots A_5}$, the latter two being totally graded
antisymmetric. The nonvanishing (anti) commutation rules of the
M-algebra are given by \cite{es1} 
\bea
\{Q_\a, Q_\b\} &=& \g^\m_{\a\b}\ P_\m + \g_{\m\a\b}\ Z^\m 
                  +\g_{\m\n\a\b}\ Z^{\m\n}+\g_{\5\a\b}\ Z^{\5}\ , \nn\\
\es
\[P_\m,Q_\a\] &=& \g_{\m\a\b}\ Z^\b -\g_{\m\n\a\b}\ Z^{\n\b}
              -\g_{\m\n_1\cdots \n_4\a\b}\ Z^{\n_1\cdots \n_4\b}\ , \nn\\
\es
\[P_\m, P_\n \] &=& \g_{\m\n\a\b}\ Z^{\a\b} 
                    + \g_{\m\n\m_1\cdots\m_3\a\b}\ Z^{\m_1\cdots\m_3\a\b}\
,\nn\\
\es
\[Q_\a, Z^\m\] &=& (1-\l-\tau)~\g^\m_{\a\b}\ Z^\b\ , \nn\\
\es
\[P_\l, Z^{\m\n}\] &=& \ft12 \d_\l^\m \left( \g^\n_{\a\b}\ Z^{\a\b}
                        -3\g_{\r\s\a\b}\ Z^{\n\r\s\a\b}\right)
                        +3\g_{\l\r\a\b}\ Z^{\m\n\r\a\b} \ , \nn\\
\es
\[Q_\a, Z^{\m\n}\] &=& -\ft{\l}{10} \g^{\m\n}_{\a\b}\ Z^\b
                        +\g^\m_{\a\b}\ Z^{\n\b} 
                        -6 \g_{\r\s\a\b}\ Z^{\m\n\r\s\b}\ ,\nn\\ 
\es
\[P_\m,Z^{\n\a}\] &=& -2\d_\m^\n\ \g_{\l\tau\b\g}\ Z^{\l\tau\a\b\g}
                     +10\g_{\m\tau\b\g}\ Z^{\n\tau\a\b\g}\ , \nn\\
\es
\{Q_\a, Z^{\m\b}\} &=& \ft14 \d_\a^\b\ \g^\m_{\g\d}\ Z^{\g\d}
                       +2 \g^\m_{\a\g}\ Z^{\b\g}
                   +\ft34 \d_\a^\b\ \g_{\n\r\g\d}\ Z^{\m\n\r\g\d}
                 +6 \g_{\n\r\a\g}\ Z^{\m\n\r\b\g}\ , \nn\\
\es
\[Q_\g, Z^{\a\b}\] &=& -5 \g_{\m\n\g\d}\ Z^{\m\n\a\b\d}
                      - \d_\g^\a\ \g_{\m\n\d\e}\ Z^{\m\n\b\d\e}\ ,\nn\\
\es
\[P_\m, Z^{\a\b}\] &=& -3\g_{\m\n\g\d}\ Z^{\n\a\b\g\d}\ , \nn\\
\es
\[Z^{\m\n}, Z^{\r\s}\] &=& -3 \g^\m_{\a\b}\ Z^{\n\r\s\a\b}\ , \nn\\
\es
\[Z^{\m\n}, Z^{\r\a}\] &=& -6\g^\m_{\b\g}\ Z^{\n\r\a\b\g}
                            +2\g^\r_{\a\b}\ Z^{\m\n\a\b\g}\ , \nn\\
\es
\[Z^{\m\n}, Z^{\a\b}\] &=& -2\g^\m_{\g\d}\ Z^{\n\a\b\g\d}\ , \nn\\
\es
\{Z^{\m\a}, Z^{\n\b}\} &=& -3\g^\m_{\g\d}\ Z^{\n\a\b\g\d}\ , \nn\\
\es
\[Z^{\m\a}, Z^{\b\g}\] &=& 6\g^\m_{\d\e}\ Z^{\a\b\g\d\e}\ ,\nn\\
\es
\[Q_\a, Z^{\5}\] &=& -\ft{\tau}{720} \g^{\5}_{\a\b}\ Z^\b 
                     +\g^{\m_5}_{\a\b}\ Z^{\4\b}\ ,\nn\\
\es
\[P_\l, Z^{\5}\] &=& \ft12 \d_\l^{\m_1}\ \g^{\m_2}_{\a\b}\ Z^{\m_3\cdots
\m_5\a\b}
                      \ , \nn\\
\es
\{Q_\a, Z^{\b\m_1\cdots\m_4}\} &=& \ft14 \d_\a^\b\ \g^{\m_1}_{\g\d}\ 
        Z^{\m_2\cdots\m_4\g\d}+2\g^{\m_1}_{\a\g}\ Z^{\m_2\cdots\m_4\b\g}\ ,\nn\\
\es
\[P_\l, Z^{\b\4}\] &=& 2\d_\l^{\m_1}\ \g^{\m_2}_{\g\d}\ Z^{\m_2\m_3\b\g\d}\
,\nn\\
\es
\[Q_\a, Z^{\m\n\r\b\g}\] &=& \d_\a^\b\ \g^\m_{\d\e}\ Z^{\n\r\g\d\e}
                             +5\g^\m_{\a\e}\ Z^{\n\r\b\g\e}\ ,\nn\\
\es
\[P_\l, Z^{\m\n\r\b\g}\] &=& -\d_\l^\m\ \g^\n_{\d\e}\ Z^{\r\b\g\d\e}\ ,\nn\\
\es
\{Q_\d, Z^{\m\n\a\b\g}\} &=& -\ft3{10} \d_\d^\a\ \g^\m_{\e\k}\ Z^{\n\b\g\e\k}
                              -\ft65 \g^\m_{\d\e}\ Z^{\n\a\b\g\e}\ ,\nn\\
\es
\[P_\l, Z^{\m\n\a\b\g}\] &=& -\ft35 \d^\m_\l\ \g^\n_{\d\e}\ Z^{\a\b\g\d\e}\
,\nn\\
\es
\[Q_\b, Z^{\m\a_1\cdots\a_4}\] &=& \d^{\a_1}_\b\ \g^\m_{\g\d}\
Z^{\a_2\cdots\a_4\g\d}+\ft72 \g^\m_{\b\g}\ Z^{\a_1\cdots\a_4\g}\ .\la{ms}
\eea
where it is understood that the obvious symmetries of indices on the
left hand side are to be implemented on the right hand side, with unit
strength (anti) symmetrizations. Thus, all free bosonic indices are to
be antisymmetrized, and all free fermionic indices are to be symmetrized
with unit strength. The parameters $\l$ and $\tau$ are arbitrary.

Several properties of the above algebra have already been discussed in
Ref. [5]. It should be stressed that this algebra is clearly not
isomorphic to $OSp(1|32)$, nor is it related to it any obvious way. A
related remark is that the M-algebra does not seem to indicate a $10+2$
dimensional origin, or hint at any obvious 10+2 dimensional extension, which we 
consider to be an interesting open problem. It is also worth mentioning that the 
the existence of the M-algebra is nontrivial, and requires that the following 
$\gamma$-matric identity to be satisfied:
\be
 \g_{\m\n(\a\b}\,\g^\n_{\g\d)} = 0\ , \la{id1}
\ee
which holds in $D=4,5,7,11$ \cite{ac}. Relevant consequences of this identy are
\bea
&&\g_{\l(\a\b} \g^{\l\m\n\r\s}_{\g\d)}-
        3 \g^{[\m\n}_{(\a\b} \g^{\r\s]}_{\g\d)}=0\ ,  \la{id2}\\
&&\g_{\m(\a\b} \g^\m_{\g\d)}+\ft1{10}\g_{\m\n(\a\b}
      \g^{\mu\nu}_{\g\d)}=0\ , \la{id3}\\
&&\g_{\m(\a\b} \g^\m_{\g\d)}+ 
       \ft1{720}\g_{\m_1\cdots \m_5 (\a\b}
            \g^{\m_1\cdots \m_5}_{\g\d)}=0\ . \la{id4}
\eea

The generators $Z^A$ decouple from the algebra if we set $\l=\tau=0$ and
redefine the translation generator as $P_\m+Z_\m \equiv P'_\m$. It is
also easy to see that the generators $Z^{A_1\cdots A_5}$ can be
Wigner-\.In\"on\"u contracted away. However, it does not seem to be
possible to Wigner-\.In\"on\"u contract the generators $Z^{AB}$ while
maintaining $Z^A$ and/or $Z^{A_1\cdots A_5}$. This can also be
understood from the fact that the $\{Q,\{Q,Q\}\}$ Jacobi identity
requires the $\gamma$- matrix identity (\ref{id2}), in which the second
term originates from the $Z^{\m\n}$ generator. 

The M-algebra can be exponentiated to describe the corresponding
supergroup manifold. Denoting collectively by $T_{\hat A}$ all the
generators of the M-algebra, and by $Z^{\hat M}$ the corresponding
supergroup manifold coordinates, we can define the basis super one-forms
as follows
\be
e^{\hat A}= dZ^{\hat M}~\left( U^{-1}\partial_{\hat M} U\right)^{\hat A}\ ,
\ee
where $U$ is an element of the supergroup, which can be parametrized in a
variety of ways \cite{es1}. 

It is useful to construct closed super-forms that may facilitate the
construction of the required Wess-Zumino terms, in search for novel
super $p$-brane actions based on the M-algebra. To this end let
us consider the following super-forms:
\bea
H_3 &=& e^\m\v e^\a\v e^\b\ \g_{\m\a\b}\ , \la{h3f}\\
H_4 &=& \ft14 e^\m\v e^\n\v e^\a\v e^\b\ \g_{\m\n\a\b}\ , \la{h4f}\\
H_7 &=& H_4\v H_3\ . \la{h7nn}
\eea
Using the Maurer-Cartan structure equtions based on the M-algebra, one
can show that these forms are indeed closed. Locally we can define 
the corresponding potentials as $H_3=dC_2$, $H_4=dC_3$ and $H_7=dC_6$,
where
\bea
C_2 &=& -e^\a\v e'_\a \ ,  \la{b2}\\
C_3 &=& -\ft16 e^\m\v e^\n\v e_{\m\n}-\ft3{20} e^\m\v e^\a\v e_{\m\a} 
        +\ft1{30} e^\a\v e^\b\v e_{\a\b} \ , \la{b3n}\\
C_6 &=& C_3\v H_3\ .\la{b6nn}
\eea

There is an alternative, and possibly more significant, example of a
closed super seven-form given by
\be
H'_7 =\ft1{5!}\ e^{\m_1}\v \cdots e^{\m_5}\v e^\a\v e^\b\ \g_{\m_1\cdots
\m_5\a\b}
      + H_4\v C_3\ ,\la{h7n}
\ee
Indeed $dH_7=0$, and writing $H_7=dC_6$, we find that $C_6$ is  given by
\cite{es1}
\bea
C'_6 =&& \ft1{5!\times 77}\, \big( -\ft{77}3\, e^{\m_1}\v \cdots \v e^{\m_5}\v
                                              e_{\m_1\cdots \m_5} 
        +\ft{281}6\, e^{\m_1}\v \cdots \v e^{\m_4}\v e^\a\v 
                     e_{\m_1\cdots \m_4\a}\nn\\
   && +\ft{104}3\, e^\m\v e^\n\v e^\r\v e^\a\v e^\b\v e_{\m\n\r\a\b}         
      -\ft{47}6\, e^\m\v e^\n\v e^\a\v e^\b\v e^\g\v e_{\m\n\a\b\g}\nn\\
   && +5\, e^\m\v e^{\a_1}\v \cdots \v e^{\a_4}\v e_{\m\a_1\cdots \a_4}
      - \ft59\, e^{\a_1}\v \cdots \v e^{\a_5}\v e_{\a_1\cdots \a_5}\nn\\
   && -\ft{131}3\, e^\m\v e^\n\v e^\r\v e^\a\v e_{\m\n}\v e_{\r\a}
        +\ft{50}3\, e^\m\v e^\n\v e^\a\v e^\b\v e_{\m\n}\v e_{\a\b}\nn\\
   && +\ft{20}3\, e^\m\v e^\a\v e^\b\v e^\g\v e_{\m\a}\v e_{\b\g}\ \big)
    \ .\la{b6}
\eea

The pull-backs of the $C$-forms to suitable worldvolumes are candidate
Wess-Zumino terms for the corresponding super $p$-branes. For a further
discussion of this point, we refer the reader to Ref. [5].

We now turn to the main point of this note, namely the chiral reductions
of the M-algebra down to ten and six dimensions. We begin by reduction
to ten dimensions. The single Majorana spinor of eleven dimensions
decomposes into a left- and a right-handed Majorana-Weyl spinor in ten
dimensions. Using a well established chiral notation, we denote the
left-handed spinors by upper spinor indices, and the right-handed ones
by lower spinor indices, and it is understood that there can be no raising
and lowering of indices. For example, the left-handed supercharges that
genarate the $(1,0)$ supersymmetry will be denoted by $Q^\a$, and the
right- handed oned that generate the $(0,1)$ supersymmetry by $Q_\a$,
where $\a=1,...,16$. 

Keeping all the generators in ten dimensions clearly yields the 
M-extension of Type IIA Poincar\'e superalgebra. Keeping only 
$Q_A, Z^A, Z^{A_1\cdots A_5}$ gives rise to the M-extension of the
$(1,0)$ Poincar\'e superalgebra that underlies the heterotic string.
This is not obvious. One way to show that the resulting algebra indeed
satisfies the Jacobi identities is to perform an elaborate
Wigner-\.In\"on\"u contraction scheme, followed by redefinion of
generators that carry the same Lorentz representations. Another approach
is to keep the aforementioned generators in the algebra and to
explicitly check the integrability of the Maurer-Cartan structure equations.
Having performed the full calculation in the case of eleven dimensional
M-algebra \cite{es1}, it turns out to be easier to follow the
latter procedure. In doing so, it is important to note that the analogs
of the $\gamma$- matrix identities in ten dimensions are
\bea
&&\g_{\l(\a\b} \g^{\l\m\n\r\s}_{\g\d)}=0\ ,  \la{id5}\\
&&\g_{\m(\a\b} \g^\m_{\g\d)}=0\ , \la{id6}\\
&&\g_{\m_1\cdots \m_5 (\a\b}
            \g^{\m_1\cdots \m_5}_{\g\d)}=0\ . \la{id7}
\eea

Thus, we find that the M-extension of the $(1,0)$ Poincar\'e superalgebra in ten
dimensions is given by:
\bea
\{Q_\a, Q_\b\} &=& \g^\m_{\a\b}\ P_\m + \g_{\m\a\b}\ Z^\m 
                 +\g_{\5\a\b}\ Z^{\5}\ , \nn\\
\es
\[P_\m,Q_\a\] &=& \g_{\m\a\b}\ Z^\b 
              -\g_{\m\n_1\cdots \n_4\a\b}\ Z^{\n_1\cdots \n_4\b}\ , \nn\\
\es
\[P_\m, P_\n \] &=& \g_{\m\n\m_1\cdots\m_3\a\b}\ Z^{\m_1\cdots\m_3\a\b}\
,\nn\\
\es
\[Q_\a, Z^\m\] &=& -\g^\m_{\a\b}\ Z^\b\ , \nn\\
\es
\[Q_\a, Z^{\5}\] &=& -\g^{\5}_{\a\b}\ Z^\b 
                     +\g^{\m_5}_{\a\b}\ Z^{\4\b}\ ,\nn\\
\es
\[P_\l, Z^{\5}\] &=& \ft12 \d_\l^{\m_1}\ \g^{\m_2}_{\a\b}\ Z^{\m_3\cdots
\m_5\a\b}
                      \ , \nn\\
\es
\{Q_\a, Z^{\b\m_1\cdots\m_4}\} &=& \ft14 \d_\a^\b\ \g^{\m_1}_{\g\d}\ 
        Z^{\m_2\cdots\m_4\g\d}+2\g^{\m_1}_{\a\g}\ Z^{\m_2\cdots\m_4\b\g}\ ,\nn\\
\es
\[P_\l, Z^{\b\4}\] &=& 2\d_\l^{\m_1}\ \g^{\m_2}_{\g\d}\ Z^{\m_2\m_3\b\g\d}\
,\nn\\
\es
\[Q_\a, Z^{\m\n\r\b\g}\] &=& \d_\a^\b\ \g^\m_{\d\e}\ Z^{\n\r\g\d\e}
                             +5\g^\m_{\a\e}\ Z^{\n\r\b\g\e}\ ,\nn\\
\es
\[P_\l, Z^{\m\n\r\b\g}\] &=& -\d_\l^\m\ \g^\n_{\d\e}\ Z^{\r\b\g\d\e}\ ,\nn\\
\es
\{Q_\d, Z^{\m\n\a\b\g}\} &=& -\ft3{10} \d_\d^\a\ \g^\m_{\e\k}\ Z^{\n\b\g\e\k}
                              -\ft65 \g^\m_{\d\e}\ Z^{\n\a\b\g\e}\ ,\nn\\
\es
\[P_\l, Z^{\m\n\a\b\g}\] &=& -\ft35 \d^\m_\l\ \g^\n_{\d\e}\ Z^{\a\b\g\d\e}\
,\nn\\
\es
\[Q_\b, Z^{\m\a_1\cdots\a_4}\] &=& \d^{\a_1}_\b\ \g^\m_{\g\d}\
Z^{\a_2\cdots\a_4\g\d}+\ft72 \g^\m_{\b\g}\ Z^{\a_1\cdots\a_4\g}\ ,\la{ha}
\eea
where $Z^{\5}$ is self-dual. The arbitrary constants $\l$ and $\tau$
that occur in \eq{ms} have now been absorbed into the definition of
$Z^\a$. This is possible because of the $\g$-matrix identities \eq{id6}
and \eq{id7}. As in the case of M-algebra, the generators $Z^A$ can
consistently be eliminated, by a redefinition of $P_\mu$ and a simple
Wigner-\.In\"on\"u contraction. $P_\m$ and $Z^{\5}$ parametrize a
symmetric $16\times 16$ matrix: $10+126=136$. As in the case of the
M-algebra, the generators $Z^\a$ and $Z^{\a_1\cdots \a_5}$ commute with
all the other generators, except Lorentz generators. The commutation
rules involving the Lorentz generators are the obvious ones, and
therefore they have not been given above. It should be stressed that the
M-algebra, as well as its ten dimensional heterotic truncation, can be
extended by the inclusion of the Lorentz generators, with the usual
commutation rules. 

For completeness, we shall also give the Maurer-Cartan structure equations, 
which can easily be read off from \eq{ha}:
\bea
de^\m &=& -\ft12~e^\a\v e^\b\ \g^\m_{\a\b}\ , \nn\\
\es
de^\a &=& 0\ , \nn\\
\es
de'_\m &=& -\ft12 e^\a\v e^\b\ \g_{\m\a\b}\ , \nn\\
\es
de'_\a &=& -e^\b\v e^\m\ \g_{\m\a\b}+\ e^\b\v e'_\m\ \g^\m_{\a\b}
         +\ e^\b\v e_{\m_1\cdots \m_5}\ 
           \g^{\m_1\cdots \m_5}_{\a\b}\ ,\nn\\
\es
de_{\5} &=& -\ft12 e^\a\v e^\b\ \g_{\5\a\b}\ , \nn\\
\es
de_{\4\a} &=& e^\b\v e^\tau\ \g_{\tau\4\a\b}+ 
              e^\b\v e_{\tau\4}\ \g^\tau_{\a\b}\ ,\nn\\
\es
de_{\m\n\r\a\b} &=& \ft12 e^\s\v e^\tau\ \g_{\s\tau\m\n\r\a\b} 
                    -\ft12 e_{\m\n\r\s\tau}\v e^\s\ \g^\tau_{\a\b}
                    +\ft14 e_{\m\n\r\tau\g}\v e^\g\ \g^\tau_{\a\b}   \nn\\
		&&-2 e_{\tau\m\n\r\a}\v e^\g\ \g^\tau_{\b\g}\ , \nn\\
\es
de_{\m\n\a\b\g} &=& 2 e^\s\v e_{\s\tau\m\n\a}\ \g^\tau_{\b\g}
                    + e^\d\v e_{\tau\m\n\d\a}\ \g^\tau_{\b\g}  
                    +5 e^\d\v e_{\tau\m\n\a\b}\ \g^\tau_{\g\d} \nn\\
\es
de_{\m\a_1\cdots\a_4} &=& -e^\n\v e_{\m\n\tau\a_1\a_2}\ \g^\tau_{\a_3\a_4}
               -\ft3{10} e^\g\v e_{\m\n\g\a_1\a_2}\ \g^\n_{\a_3\a_4} \nn\\
               &&-\ft65 e^\g\v e_{\m\n\a_1\a_2\a_3}\ \g^\n_{\a_4\g} \nn\\
\es
de_{\a_1\cdots\a_5} &=& -\ft35 e^\m\v e_{\m\n\a_1\a_2\a_3}\ \g^\n_{\a_4\a_5} 
                       + e^\g\v e_{\m\g\a_1\a_2\a_3}\ \g^\m_{\a_4\a_5}\nn\\
              &&+\ft72 e^\g\v e_{\m\a_1\cdots \a_4}\ \g^\m_{\a_5\g}\ , \la{malg}
\eea
where, as before, it is understood that the obvious symmetries of indices on the
left hand side are to be implemented on the right hand side, with unit
strength (anti) symmetrizations. The freedom in choosing the coefficents
of the $Z^\a$ generators in the algebra \eq{ha} translates into the
freedom of choice for all the coefficients on the right hand side
of the equation for $de^{\a'}$. 

The closed super-forms associated with the algebra \eq{ha} that can be
used in the constrcution of te Wess-Zumino terms for strings and
fivebranes in ten dimensions are:
\bea
H_3 &=& e^\m\v e^\a\v e^\b\ \g_{\m\a\b}
+e^\a\v e^\b \left(e'_\m\ \g^\m_{\a\b}
    +e_{\5}\ \g^{\5}_{\a\b}\right)\ ,\la{h3nn}\\
H_7 &=& \ft1{5!} e^{\m_1}\v \cdots \v e^{\m_5}\v e^\a\v e^\b\ 
              \g_{\m_1\cdots \m_5\a\b}\ . \la{h7ff}
\eea
The associated potentials defined by $H_3=dC_2$ and $H_7=dC_6$ are given
by
\bea
C_2 &=& -e^\a\v e'_\a \ ,  \la{b2n}\\
C_6 &=& \ft1{5!\times 77}\, \big( -\ft{77}3\, e^{\m_1}\v \cdots \v e^{\m_5}\v
                                              e_{\m_1\cdots \m_5} 
        +\ft{281}6\, e^{\m_1}\v \cdots \v e^{\m_4}\v e^\a\v 
                     e_{\m_1\cdots \m_4\a}\nn\\
   && +\ft{104}3\, e^\m\v e^\n\v e^\r\v e^\a\v e^\b\v e_{\m\n\r\a\b}         
      -\ft{47}6\, e^\m\v e^\n\v e^\a\v e^\b\v e^\g\v e_{\m\n\a\b\g}\nn\\
   && +5\, e^\m\v e^{\a_1}\v \cdots \v e^{\a_4}\v e_{\m\a_1\cdots \a_4}
      - \ft59\, e^{\a_1}\v \cdots \v e^{\a_5}\v e_{\a_1\cdots \a_5}\ \big)
    \ .\la{b6n}
\eea
It is clear that every term in \eq{h3nn} is separately closed. Setting
the last two terms equal to zero implies that the pull-back of $C_2$,
which can serve as a Wess-Zumino term, contains the new coordinates
associated with $Z^\m$ and $Z^{\m_1\cdots \m_5}$ only in a total
derivative term. Adding a standard kinetic term, this would then
furnish a reformulation of the heterotic string. However, keeping the
second or the third term in \eq{h3nn} implies that the new coordinates
are no longer contained in a total derivative term. New kinds of
kinetic terms and new bosonic as well as fermionic symmetries would be
needed in that case, to construct a novel string action.

In the case of superfivebrane, \eq{h7ff} shows that the dependence on
all the new coordinates are indeed confined to a total derivative, and
hence, the pull-back of \eq{b6n}, together with a standard kinetic term 
for the ordinary superspace coordinates, provides a reformulation of the 
usual superfivebrane action \cite{bst}.

We now turn to the reduction down to six dimensions. Starting from
eleven dimensions, the Lorentz group $SO(10,1)$ decomposes into
$SO(5,1)\times Sp(2)$, where $Sp(2) \sim SO(5)$ is the diagonal subgroup
of the automorphism group $Sp(2)_I\times Sp(2)_{II}$ of the $(2,2)$
Poincar\'e superalgebra in six dimensions. The 32 component supercharges
of eleven dimensions decompose into a left-handed symplectic Majorana-Weyl
charge $Q_{\a i}$, and a right-handed one $Q^\a_i$, with $\a, i=1,...,4$.
The translation generators of eleven dimensions decompose into $P_\mu$ and
$P_{a'}$ with $\mu=0,1,...,5$ and $a'=1,...,5$. Decomposing all the
generators of the M-algebra accordingly, and keeping all of them, one
obtains the M-extension of the $(2,2)$ Poincar\'e superalgebra in six
dimensions. A chiral truncation of this algebra to $(2,0)$ supersymmetry can be 
achieved by keeping the following generators
\be
\noindent (2,0)~~{\rm Algebra~ Generators:}\quad\quad 
  Q_A\ ,~~Z^A\ ,~~Z^{Aa'}\ ,~~Z^{ABCa'b'}\ , \la{20}
\ee
where $Q_A=(Q_{\a i}, P_\m)$. The $\{Q,Q\}$ anticommutator produces
$P_\m$, five vector charges $Z^{\m a'}$ and ten self-dual three-form
charges $Z^{\m\n\r a'b'}$, which parametrize a $16\times 16$
symmetric matrix that has $136$ components: $6+5\times 6+10\times
10=136$. The vectorial generator $Z^\m$ is also produced, but it is
really not an independent generator, since it can be absorbed into the
definition of $P_\m$. Nonetheless, it is expected to play a significant
role in the description of strings with winding states \cite{pkt2}. The
remaining generators of the M-algebra with even number of Lorentz
indices such as $P_{a'}$, $Z^{\m\n}$, $Z^{\m\n\r\s a'}$, ... are
projected out by chirality, and those whith an odd number of vector
indices are either projected out by duality/chirality properties of the
$\g$-matrices or they can be redefined away as part of the generators
already mentioned. 

In order to verify that the generators \eq{20} form a closed algebra,
and that the remaining generators of the M-algebra decouple, one has
to show that a suitable \wi~contraction scheme exists. We have shown
that this is indeed the case, by scaling the generators with suitable
powers of a parameter and then taking the limit where this parameter
vanishes. It is straightforward to read off the (anti) commutation rules
of the resulting algebra from \eq{ms}.

It is also possible to reduce the $(2,2)$ algebra described earlier to $(1,1)$ and
$(1,0)$ algebras. The easiest way to do this is to start from \eq{ha}
which represents the chirally reduced version of the M-algebra in ten
dimensions. We can proceed exactly as in the case of $11 \rightarrow 6$ reduction
described above. Essentially one replaces $Sp(2)$ by $Sp(1)$ in the
above discussion. To be more specific, the Lorentz group $SO(9,1)$
decomposes into $SO(5,1)\times SO(4)$, where $SO(4) \sim Sp(1)_I\times
Sp(1)_{II}$ is the automorphism group of the $(1,1)$ Poincar\'e
superalgebra in six dimensions. The 16 component supercharges of ten
dimensions decompose into a left-handed symplectic Majorana-Weyl charge
$Q_{\a i}$, and a right-handed one $Q^\a_i$, with $\a=1,...,4$, $i=1,2$. The
translation generators of ten dimensions decompose into $P_\mu$ and
$P_{a'}$ with $\mu=0,1,...,5$ and $a'=1,...,4$. Decomposing all the
generators of the M-algebra accordingly, and keeping all of them, one
obtains the M-extension of the $(1,1)$ Poincar\'e superalgebra in six
dimensions. A chiral truncation of this algebra to $(1,0)$ supersymmetry
can be achieved by keeping the following generators 
\be 
\noindent (1,0)~~{\rm Algebra~ Generators:}\quad\quad 
Q_A\ ,~~Z^A\ ,~~Z^{ABCa'b'}\ . \la{10} 
\ee 
The $\{Q,Q\}$ anticommutator produces $P_\m$, and a triplet of
three-form charges $Z^{\m\n\r a'b'}$, which parametrize an $8\times 8$
symmetric matrix that has $36$ components: $6+3\times 10=36$. The
generators $Z^\m$ are also produced, but they can be absorbed into the
definition of $P_\m$, as discussed earlier, at least locally. Note that
$Z^{\m\n\r a'b'}$ is self-dual in $SO(5,1)$, and in $SO(4)$. All the
generators of the algebra \eq{ha} that are outside the set \eq{10} can
be truncated by arguments parallel to those given above for the case of
$(2,2) \rightarrow (2,0)$ reduction. Note, in particular, that the 
super two-form generators drop out. This procedure yields the following
M-extension of the $(1,0)$ Poincar\'e superalgebra:
\bea
\{Q_\a, Q_\b\} &=& \g^\m_{\a\b}\ P_\m + \g_{\m\a\b}\ Z^\m 
                 +10 \left(\g_{\m\n\r a'b'}\right)_{\a\b}\ Z^{\m\n\r a'b'}\ , \nn\\
\es
\[P_\m,Q_\a\] &=& \g_{\m\a\b}\ Z^\b 
              -6 \left(\g_{\m\n\r a'b'}\right)_{\a\b}\ Z^{\n\r\b a'b'}\ , \nn\\
\es
\[P_\m, P_\n \] &=& 3\left(\g_{\m\n\r a'b'}\right)_{\a\b}\ Z^{\r\a\b a'b'}\
,\nn\\
\es
\[Q_\a, Z^\m\] &=& -\g^\m_{\a\b}\ Z^\b\ , \nn\\
\es
\[Q_\a, Z^{\m\n\r a'b'}\] &=& -\g^{\5}_{\a\b}\ Z^\b 
                     +\ft35 \g^\m_{\a\b}\ Z^{\n\r\b a'b'}\ ,\nn\\
\es
\[P_\l,  Z^{\m\n\r a'b'}\] &=& \ft3{20} \d_\l^\m\ \g^\n_{\a\b}\ Z^{\r\a\b a'b'}
                      \ , \nn\\
\es
\{Q_\a, Z^{\m\n\b a'b'}\} &=& \ft18 \d_\a^\b\ \g^\m_{\g\d}\ 
        Z^{\n\g\d a'b'}+ \g^\m_{\a\g}\ Z^{\n\b\g a'b'}\ ,\nn\\
\es
\[P_\l,  Z^{\m\n\b a'b'}\] &=& \d_\l^\m\ \g^\n_{\g\d}\ Z^{\b\g\d a'b'}\
,\nn\\
\es
\[Q_\a, Z^{\m\b\g a'b'}\] &=& \ft13 \d_\a^\b\ \g^\m_{\d\e}\ Z^{\g\d\e a'b'}
                             +\ft53 \g^\m_{\a\d}\ Z^{\b\g\d a'b'}\ ,
\la{6da}
\eea
with the symmetries of the left hand side indices to be implemented on the right
hand side, with unit strength, as usual. We have used a condensed notation in which the
index $\a$ stands for the pair $\a i$ with $\a=1,...,4$, $i=1,2$. As for
the $\g$-matrices, the following replacements are to be made:
$\g^\m_{\a\b} \rightarrow \g^\m_{\a\b}~\e_{ij}$ and $\g^{\m\n\r
a'b'}_{\a\b} \rightarrow \g^{\m\n\r}_{\a\b}~\g^{a'b'}_{ij}$, where
$\e_{ij}$ is the $Sp(1)$ invariant tensor, and $\g^{a'b'}_{ij}$ is the
self-dual projection of the $SO(4)$ generators. 

To conclude, we have obtained the chiral reductions of the M-algebra
down to the $(1,0)$ algebra in ten and the $(2,0)$ and $(1,0)$ algebras
in six dimensions. These algebras contain various super $p$-form
charges. The subset of these charges which arise in the anticommutator
of the supercharges $Q_\a$ are known to be associated with super
$p$-brane solitons that arise in the corresponding supergravity
theories. It would be interesting to see if novel soliton solutions of
six dimensional supergravities exist in which all
the bosonic three-form charges that occur in the above algebra are
activated. It is concievable that they arise in certain compactifications of
M-theory down to six dimensions. 

The significance of the charges other than the supercharge $Q_\a$ that carry
spinor indices is less clear at present. In the case of strings, the
spinorial charge $Z^\a$ has been utilized \cite{g}
in finding a Chern-Simons origin of the Green-Schwarz superstring
action. It has also been used \cite{s1} in a
reformulation of the superstring which can be put on a
lattice. An analogous reformulation of the supermembrane based on a
truncated version of the M-algebra has already been found \cite{bs2}.
However, the role of the full M-algebra in a formulation of superfivebrane 
theory is yet to be discovered.

There are a number of intersting questions that are left open: Is there
any connection between the M-algebra and the classified simple Lie
superalgebras or their contractions thereof? Can the M-algebra be
extended to $10+2$ dimensions? Is there a membrane-fivebrane duality
symmetry in M-theory that can be understood at the level of the
M-algebra? Is there a formulation of M-branes that makes a nontrivial
use of the new coordinates associated with the new generators of the
M-algebra, and if so, can these coordinates play a role in exploring
the duality symmetries of M-theory? What is the ultimate role of the
M-algebra in M-theory? 

\bigskip

\noindent{\bf Acknowledgements}

\bigskip
 
This work was supported in part by NSF Grant PHY-9411543.

\vfill\eject

\end{document}